\documentclass[aps,prl,10pt,twocolumn,amsmath,amssymb,floatfix]{revtex4}

\usepackage{color}
\usepackage{graphicx}
\usepackage{txfonts}
\usepackage{braket}
\usepackage{dsfont}
\usepackage[colorlinks,citecolor=blue,linkcolor=blue,urlcolor=blue,plainpages=false,pdfpagelabels]{hyperref}

\renewcommand{\vec}[1]{{\mathbf{#1}}}
\newcommand{\bsigma}{\boldsymbol{\sigma}}
\newcommand{\pd}{{\phantom{\dagger}}}
\newcommand{\tphi}{\tilde{\phi}}

\renewcommand{\Im}{\text{Im}}
\renewcommand{\Re}{\text{Re}}
\newcommand{\Tr}{\text{Tr}}

\begin{document}1

\title{Universal Hall conductance scaling in non-Hermitian Chern insulators}

\author{Solofo Groenendijk}
\affiliation{Department of Physics and Materials Science, University of Luxembourg, L-1511 Luxembourg}

\author{Thomas~L.~Schmidt}
\affiliation{Department of Physics and Materials Science, University of Luxembourg, L-1511 Luxembourg}

\author{Tobias Meng}
\affiliation{Institute for Theoretical Physics and W\"urzburg-Dresden Cluster of Excellence ct.qmat,\\ Technische Universit\"at Dresden, 01069 Dresden, Germany}

\date{\today}

\begin{abstract}
We investigate the Hall conductance of a two-dimensional Chern insulator coupled to an environment causing gain and loss. Introducing a biorthogonal linear response theory, we show that sufficiently strong gain and loss lead to a characteristic non-analytical contribution to the Hall conductance. Near its onset, this contribution exhibits a universal power-law with a power 3/2 as a function of Dirac mass, chemical potential and gain strength. Our results pave the way for the study of non-Hermitian topology in electronic transport experiments.
\end{abstract}

\maketitle
Topology is a universal organizing principle complementing Landau's theory of symmetry-broken phases~\cite{Kosterlitz1973,Wen2017,Chiu2016}. In Landau's theory, different phases are distinguished by local order parameters. In a similar vein, topologically distinct phases can be discerned by topological invariants. Unlike local order parameters, which take continuous values, topological invariants are quantized~\cite{Wen2017,Qi2011}: they can only change in discrete steps, and often take values in $\mathds{Z}_2$ or $\mathds{Z}$~\cite{Wen2017,Kitaev2009}. In theory, topological phases can be identified by discrete outcomes of non-local measurements reflecting the topological invariants~\cite{Wen2017,Qi2011}. Prototypical examples of topological states are the integer quantum Hall state in a two-dimensional electron system~\cite{Thouless1982}, and the closely related two-dimensional Chern insulators~\cite{Haldane1988,Liu2016,Qi2011}. The topological invariant of these states is the (first) Chern number $C\in\mathds{Z}$. In principle, the Chern number can be experimentally accessed via the Hall conductance $\sigma_{xy}=(e^2/h)\,C$~\cite{Thouless1982,Liu2016}.

The strict mathematical definition of topological invariants typically requires isolated systems of infinite size~\cite{Gulden2016}. As experimental systems with non-trivial topology are neither infinite nor completely decoupled from their environment, it is important to quantify the extent to which they exhibit quantized responses at all. While finite system sizes only lead to exponentially small corrections to topologically quantized response functions~\cite{Zhou2008,Albrecht2016}, the impact of an environment and thus a coupling to external degrees of freedom is still an important open problem.

In this Letter, we study the prototypical case of a Chern insulator coupled to an environment, and consider the system in a regime where it can be described by an effective non-Hermitian Hamiltonian with loss and gain terms \cite{Giusteri2015,Bergholtz2019,Torres2019,supplement}. Whereas the amplitude of the Hall conductance in such a non-Hermitian Chern insulator ceases to be quantized \cite{Philip2018,Hirsbrunner2019}, we find that sufficiently strong loss and gain leads to a new type of universality in the Hall response: a non-analytic dependence of the Hall conductance on the variation of all relevant system parameters with the exponent $3/2$.

This behavior of the Hall conductance is somewhat similar to the behavior of the density of states in systems with loss and gain exhibiting so-called exceptional points. At these points in parameter space, several of the eigenvectors of the effective single-particle Hamiltonian coalesce, such that they no longer form a complete basis~\cite{Berry2004,Rotter2009,Shen2018,Torres2019,Bergholtz2019}. In an electronic measurement, exceptional points are reflected in a characteristic non-analytic contribution to the density of states scaling with a universal exponent $1/2$ \cite{Papaj2019}. The non-analytic contribution to the Hall conductance identified in this work has an equally universal scaling, albeit with the exponent 3/2. While it can appear in a similar parameter regime as required for exceptional points, we show that both phenomena in general occur independently from one another. Our findings thus not only provide a new quantized fingerprint of non-Hermitian physics, but they do so in a transport quantity that is characteristic of a topological system with otherwise non-quantized response functions, namely the non-Hermitian Chern insulator.

We begin by presenting a generic theory of open electronic systems. The starting point is a Hamiltonian describing both the system and its environment. In many cases, the effect of the environment can be taken into account by introducing non-Hermitian terms in an effective system Hamiltonian. Non-Hermitian Hamiltonians already appear at the level of retarded Green's functions with complex self-energies, $G^R(\vec{k},\omega) = [\omega-\mathcal{H}_0(\vec{k})+\Sigma^R(\vec{k},\omega)]^{-1}$, where $\mathcal{H}_0$ is the Bloch Hamiltonian of the isolated system and $\Sigma^R$ is the retarded self-energy brought about by its coupling to the environment. This can in turn be used to construct an effective non-Hermitian Bloch Hamiltonian $\mathcal{H}_{\rm nh}(\vec{k}) := \mathcal{H}_0(\vec{k})-\Sigma^R(\vec{k},0)$. Another avenue for non-Hermitian Hamiltonians is the Lindblad Master equation $\partial_t\rho=i[\rho,H]+\sum_n(L_n^\pd\rho L_n^\dagger-\{L_n^\dagger L_n^\pd,\rho\}/2)$ that allows one to define a non-Hermitian Hamiltonian as $H_{\rm nh} = H-(i/2)\sum_n L_n^\dagger L_n^\pd$. Here, $\rho(t)$ denotes the system density matrix and $L_n$ are Lindblad operators describing quantum jumps due to the coupling to the environment. Quite generally, the dynamics of an open system can thus be described by different forms of effective non-Hermitian Hamiltonians with loss and gain terms, at least in suitable approximations \cite{Diehl2011,El-Ganainy2018,Bergholtz2019,Song2019,Naghiloo2019,Minganti2019,Tripathi2020,supplement}.

In contrast to Hermitian Hamiltonians, non-Hermitian Hamiltonians $H_{\rm nh}$ have distinct right and left eigenvectors for the same eigenvalue, i.e., $H_{\rm nh} \ket{\phi_m} = \epsilon_m \ket{\phi_m}$ and $\bra{\tphi_m}H_{\rm nh}=\bra{\tphi_m} \epsilon_m$ with $\ket{\phi_m}\neq \ket{\tphi_m}$. Furthermore, right and left eigenvectors are not necessarily mutually orthogonal, i.e., it is possible that $\braket{\phi_m|\phi_n} \neq 0$ and $\braket{\tphi_m|\tphi_n}\neq 0$ for $m \neq n$. Instead, the algebra known from Hermitian systems carries over to natural combinations of right and left eigenstates. It is for instance possible to choose the eigenstates in such a way that $\braket{\tphi_m|\phi_n}=\delta_{mn}$ and $\mathds{1}=\sum_m  \ket{\phi_m}\bra{\tphi_m}$ \cite{Brody2013}. This immediately raises the question of how objects as basic as expectation values $\braket{\mathcal{O}}$ of a system operator $\mathcal{O}$ should be interpreted. Whether the expectation value is taken with respect to right eigenstates only, left eigenstates only, or a combination of right and left eigenstates depends on the situation that is modelled by the non-Hermitian Hamiltonian. Technically, this question can be condensed into the equation of motion for the density matrix $\rho$ describing the system \cite{Herviou2019}. Setups described by a Lindblad Master equation have short-term dynamics described by only right eigenvectors, while non-Hermitian descriptions of systems with finite lifetimes rather correspond to combining right and left eigenstates in what is known as a ``biorthogonal'' quantum description.

In this Letter, we study electronic transport in a 2D electronic system with loss and gain. To describe transport in such an open setup, we consider the system coupled to an electric field. Assuming the field to be weak, we study the time evolution of the biorthogonal expectation value of electric current in linear response. Concretely, we analyze a non-Hermitian Chern insulator modelled by the effective Hamiltonian $H_{\rm nh}= \sum_{\vec{k}}\Psi^\dagger(\vec{k})  \mathcal{H}_{\rm nh}(\vec{k}) \Psi^\pd(\vec{k})$ with the non-Hermitian Bloch Hamiltonian
\begin{align}
\mathcal{H}_{\rm nh}(\vec{k}) &= k_x\,\sigma_x+k_y\,\sigma_y+m\sigma_z-\mu\sigma_0  \nonumber\\
&- i\Gamma_0 \,\sigma_0-i\left(\Gamma_x\,\sigma_x+\Gamma_y\,\sigma_y+\Gamma_z\,\sigma_z\right).\label{eq:ourHam}
\end{align}
Here, $\Psi^\dagger(\vec{k})=[c_{\uparrow}^\dagger(\vec{k}),c_{\downarrow}^\dagger(\vec{k})]$ denotes the spinor of creation operators for electrons with spin $\sigma=\uparrow,\downarrow$ and momentum $\vec{k} = (k_x,k_y)^T$. In the Bloch Hamiltonian, $\sigma_{x,y,z}$ are Pauli matrices, and $\sigma_0$ is the identity matrix. Moreover, $m$ is the Dirac mass and $\mu$ the chemical potential. Loss and gain resulting from the coupling to the environment are encoded in the non-Hermitian terms $\Gamma_{0,x,y,z}$ which we approximate as $\vec{k}$-independent. Without loss of generality, we will assume $\Gamma_{0,x,y,z} \geq 0$. The system experiences net loss if $ \Gamma_0> |\vec{\Gamma}|$, whereas one mode has net gain for $ \Gamma_0< |\vec{\Gamma}|$ \cite{supplement}.

A well-known hallmark of non-Hermitian physics is the presence of exceptional points in the complex band structure. The Hamiltonian~\eqref{eq:ourHam} exhibits exceptional points at momenta satisfying $\sqrt{(k_x-i\Gamma_x)^2+(k_y-i\Gamma_y)^2+(m-i\Gamma_z)^2}=0$. Recently, similar exceptional points have gained a lot of attention in photonic systems with parity-time (PT) symmetry  \cite{Regensburger2012,Zhen2015,Hahn2016,Oezdemir2019,Miri2019}. In these systems, the complex refractive index of the material plays a similar role as the self-energy in electronic systems. One salient feature of non-Hermitian photonic systems is that close to the exceptional points, PT symmetry breaks down and the system exhibits rather peculiar transport properties such as reflectionless unidirectional wave propagation \cite{Oezdemir2019,Miri2019,Huang2017}, loss induced transparency \cite{Guo2009}, and non-analytic frequency response in terms of the system parameter \cite{Hodaei2017}. In contrast, the transport properties of electronic systems combining strong loss and gain with non-trivial topology have not yet been studied in great detail.

The characteristic experimental observable of Chern insulators is their Hall conductance, which is quantized in the absence of coupling to an environment. To analyze the fate of the Hall conductance in the presence of loss and gain, we first define the electric current as $\vec{j} = \partial_{\vec{k}} \mathcal{H}_{\rm nh}(\vec{k})= \partial_{\vec{k}} \mathcal{H}_{0}(\vec{k})$, where $\mathcal{H}_0$ is the unperturbed part of the Bloch Hamiltonian (i.e., the Hamiltonian at vanishing electric field). We then analyze the time evolution of its expectation value $\braket{\vec{j}}(t) = \langle\tphi_0(t)|\vec{j}|\phi_0(t)\rangle$, where the initial state $\ket{\phi_0}$ is the right eigenstate $|\phi_m\rangle$ with maximal norm of $e^ {-\beta \epsilon_m}$, whereas $\ket{\tphi_0}$ denotes the corresponding left state ($\beta$ is the inverse temperature). Their time evolution follows from the generalized Schr\"odinger equation for right and left states, $i\partial_t \ket{\phi} = H_{\rm nh}\ket{\phi}$ and $i\partial_t \ket{\tphi} = H_{\rm nh}^\dagger\ket{\tphi}$. Since we allow the non-Hermitian Hamiltonian to describe both loss and gain, we cannot study the time evolution as usual assuming that the electric field was switched on at a time $t_0\to-\infty$, but instead need to keep $t_0$ finite. Using a  Lehmann spectral representation, the current $\braket{j_x}$ in $x$ direction at time $t>t_0$ due to an electric field $\vec{E} = E_{0,y} e^{-i\omega_0 t} \vec{e}_y$ with frequency $\omega_0$ in $y$ direction can then in the zero temperature limit $\beta\to\infty$ be written as (see Supplemental Material~\cite{supplement} for details)
\begin{align}\label{eq:jx}
    \braket{j_x}(t)
&=
    \frac{E_{0,y}}{i\omega_0} e^{-i \omega_0 t} \sum_{m,n}  \frac{1}{Z} \bra{\tphi_m} j_x  \ket{\phi_n} \bra{\tphi_n} j_y \ket{\phi_m}  \notag \\
&\times
    \frac{e^{ -\beta \epsilon_m}- e^{ -\beta \epsilon_n}}{\omega_0 + \epsilon_m - \epsilon_n} \left( 1 - e^{ i (\omega_0 + \epsilon_m - \epsilon_n) (t-t_0)} \right),
\end{align}
where the sums are over all eigenstates of the Hamiltonian $\mathcal{H}_{\rm nh}$, $\beta$ is the inverse temperature and $Z = \sum_{n} e^{ -\beta \epsilon_n}$ is the generalized partition function. The overall behavior of the current, in particular a non-analytic dependence on system parameters, can be understood as being handed down to Eq.~\eqref{eq:jx} from the effective conductance
\begin{align}\label{eq:sigmaxy}
    \sigma_{xy}
&=
    \frac{1}{i\omega_0}  \sum_{m,n}  \frac{1}{Z} \bra{\tphi_m} j_x  \ket{\phi_n} \bra{\tphi_n} j_y \ket{\phi_m}
    \frac{e^{ -\beta \epsilon_m}- e^{ -\beta \epsilon_n}}{\omega_0 + \epsilon_m - \epsilon_n}.
\end{align}
To evaluate this expression, we now proceed by analytic continuation to the complex plane. We start from the current-current correlation function in imaginary time (see  Supplemental Material~\cite{supplement})
\begin{align}
    - \Tr' \left[ e^{-\beta H_{\rm nh}} T_\tau j_x(\tau) j_y(\tau') \right],
\end{align}
where the trace of an operator $\mathcal{O}$ is defined with biorthogonal states as $\Tr' \mathcal{O} = \sum_n \braket{\tphi_n|\mathcal{O}|\phi_n}$. Using a Lehmann representation, it can easily be shown that the Fourier transform of this to Matsubara frequencies exactly corresponds to the conductance $\sigma_{xy}$ identified in biorthogonal perturbation theory. This allows us to formally evaluate the biorthogonal linear response in the familiar language of Matsubara Green's functions $\mathcal{G}$, for which the current takes the form $\vec{j} = - \partial_{\vec{k}} \mathcal{G}^{-1}$. The full definition of our analytic continuation scheme also includes the complex form of the loss and gain parameters $\Gamma_i$. For the case of quasiparticle lifetimes encoded in complex self-energies, the loss parameter has the Matsubara expression $\Gamma_i(\omega_n) =\Gamma_i\text{sgn}(\omega_n)$. To generalize the framework of Matsubara-based linear-response theory to describe both quasiparticle loss and gain, we include a similar sign function in our continuation of the non-Hermitian Hamiltonian to the complex plane. It is then straightforward to show that the Hall response takes the usual form \cite{Ishikawa1986,Ishikawa1987,Imai1990,Wang2010,supplement}
\begin{align}\label{eq:ishikawa}
\sigma_{xy} = \frac{e^2}{h}\frac{\varepsilon^{\mu\nu\lambda}}{24\pi^2}\int d^3q\Tr\left[(\partial_\mu \mathcal{G}^{-1})\mathcal{G}(\partial_\nu \mathcal{G}^{-1})\mathcal{G}(\partial_\lambda \mathcal{G}^{-1})\mathcal{G}\right],
\end{align}
where $q=(k_x,k_y,i\omega_n)$ is the three-momentum including the Matsubara frequency $\omega_n$ and the trace is over the resulting $2\times 2$ matrix.

The quantization of the Hall conductance in the special case of a Hermitian Chern insulator follows quite elegantly from the interpretation of Eq.~\eqref{eq:ishikawa} as a topological winding number. Namely, if $\mathcal{G}(q)$ is a smooth function of $q$, then the integral in Eq.~(\ref{eq:ishikawa}) yields the winding number of the Green's function in the space of $2\times 2$ matrices $GL(\mathbb{C},2)$ as $q$ is varied  \cite{Ishikawa1987,Tsvelik2003,Altland2010}. This integral is also referred to as Pontryagin index in the literature, and appears in the study of Yang-Mills instantons \cite{Belavin1975}, Wess-Zumino-Witten field theories \cite{Tsvelik2003,Altland2010}, and Chern-Simons theories \cite{Chern1974}. In contrast, if the Green's function $\mathcal{G}(q)$ is not continuous, the Hall response is generically non-quantized because the integrand in Eq.~\eqref{eq:ishikawa} is not smooth \cite{Chen2018,Shen2018,Hirsbrunner2019}. The quantization of the Hall conductance is already spoiled in the simple case of a Matsubara self-energy $\Sigma^M(k_x,k_y,i\omega_n)=i\,\text{sgn}(\omega_n)\,\Sigma_0 \mathds{1}_2$, describing for example the coupling to an itinerant magnet \cite{Philip2018,Hirsbrunner2019}, because this self-energy entails a discontinuous jump of $\mathcal{G}(q)$ across the real axis.

We now show that while the amplitude of the Hall response of a non-Hermitian Chern insulator is non-universal and non-quantized, universal behavior of the Hall conductance with a quantized \emph{scaling} can still arise if the non-Hermitian part of the Hamiltonian has a nontrivial matrix structure as in Eq.~\eqref{eq:ourHam}, provided one of the two modes exhibits gain.  In the following, we focus on gapped systems with $m \neq 0$ and on chemical potentials near the Dirac point, $\mu^2\leq m^2+\Gamma_0^2$. The general method to compute the Hall conductance of Chern insulators described by Eq.~(\ref{eq:ourHam}) is given in the Supplemental Material~\cite{supplement}.

If we consider the case $\Gamma_{x,y} = 0$ but allow for nonzero $\Gamma_z$ and $\Gamma_0$, the Hall conductance is the sum of a Fermi sea contribution $\sigma^{(\rm sea)}_{xy}(\Gamma_0,\Gamma_z)$ and a Fermi surface contribution $\sigma^{(\rm surface)}_{xy}(\Gamma_0,\Gamma_z)$ [see Eqs.~(\ref{eq:Fsea}) and (\ref{eq:Fsur}) in the Supplemental Material]. The former is quantized in the limit of vanishing net dissipation ($\Gamma_0 = 0$) as long as the chemical potential is in the gap ($|\mu| < |m|$).

The situation changes dramatically if either $\Gamma_x$ or $\Gamma_y$ is sufficiently large. Since the system has rotational symmetry in the $x-y$ plane, we focus on the case $\Gamma_y \neq 0$. For simplicity, we chose $\Gamma_z = 0$ for this discussion. The conductance can be computed using Eq.~(\ref{eq:ishikawa}) and one finds that the Fermi sea contribution consists of two terms [see Eqs.~(\ref{eq:Fseatotal})  and (\ref{eq:Fseatilde}) in the Supplemental Material], such that
\begin{equation}
\sigma_{xy}(\Gamma_0, \Gamma_y) =  \tilde{\sigma}^{(\rm sea)}_{xy}(\Gamma_0)+ \sigma^{(\rm surface)}_{xy}(\Gamma_0)+\sigma^{(\rm 3/2)}_{xy}(\Gamma_0,\Gamma_y).
\end{equation}
The key player in our discussion is the new contribution
\begin{align}\label{eq:contrib}
\sigma^{(3/2)}_{xy}  &= \frac{e^2}{h}\frac{m}{2\pi}\int_{-\infty}^\infty dk_x \int_{\Gamma_0}^\infty d \omega \Re \frac{\Theta\left[\mathcal{A}(k_x,\omega)\right]}{[k_x^2+m^2+(i\mu+\omega)^2]^{3/2}}, \notag \\
\mathcal{A}(k_x,\omega) &= \Gamma_y^2+\mu^2-m^2-\frac{\mu^2 \omega^2}{\Gamma_y^2}-\omega^2-k_x^2.
\end{align}
Here, the Heaviside function $\Theta$ originates from using the residue theorem when extending the integral over $k_y$ to the upper complex plane. It constrains the new contribution to the Hall conductance to appear only if
\begin{align}
\epsilon := \frac{\Gamma_y}{\Gamma_0}\,\sqrt{\frac{\Gamma_y^2+\mu^2-m^2}{\Gamma_y^2+\mu^2}}> 1.\label{eq:condition}
\end{align}
This condition can only be satisfied for $\Gamma_y > \Gamma_0$, and thus for a system with net gain in one mode. For $\mu=\Gamma_0=0$, we find that $\sigma^{(3/2)}_{xy}$ is linked to the appearance of exceptional points, since both appear for $\Gamma_y^2\geq m^2$. In general, however, exceptional points in the complex spectrum and the new contribution to the Hall conductance appear independently from one another. In particular, it is possible to have exceptional points while $\sigma^{(3/2)}_{xy}=0$ if $\Gamma_0 \geq \Gamma_y$, or to have no exceptional points but nonzero $\sigma^{(3/2)}_{xy}$ if $\mu \neq 0$ and $\Gamma_y \gg \Gamma_0$.

\begin{figure}
\centering
\begin{tabular}{cc}
\includegraphics[width=4cm]{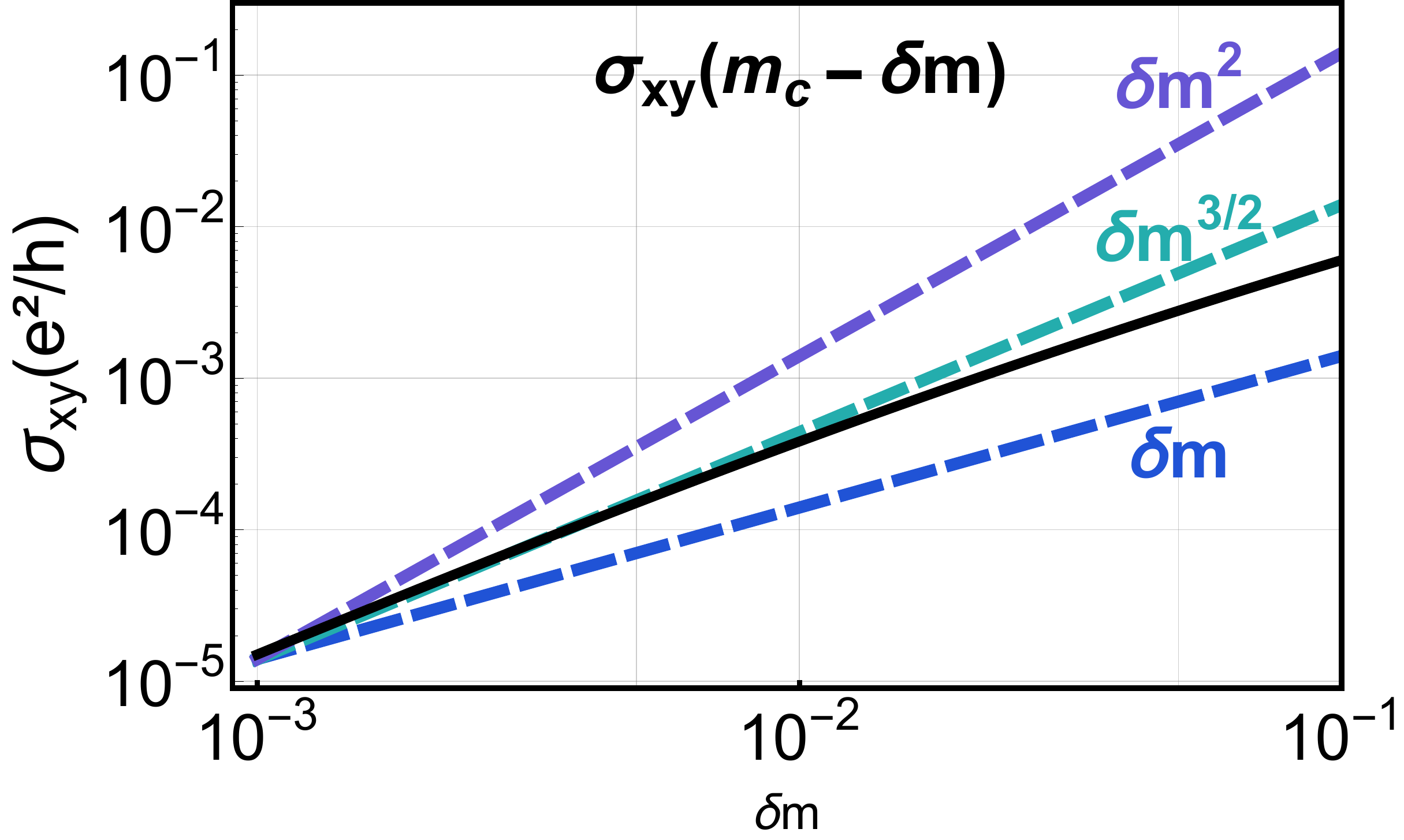} &
\includegraphics[width=3.4cm]{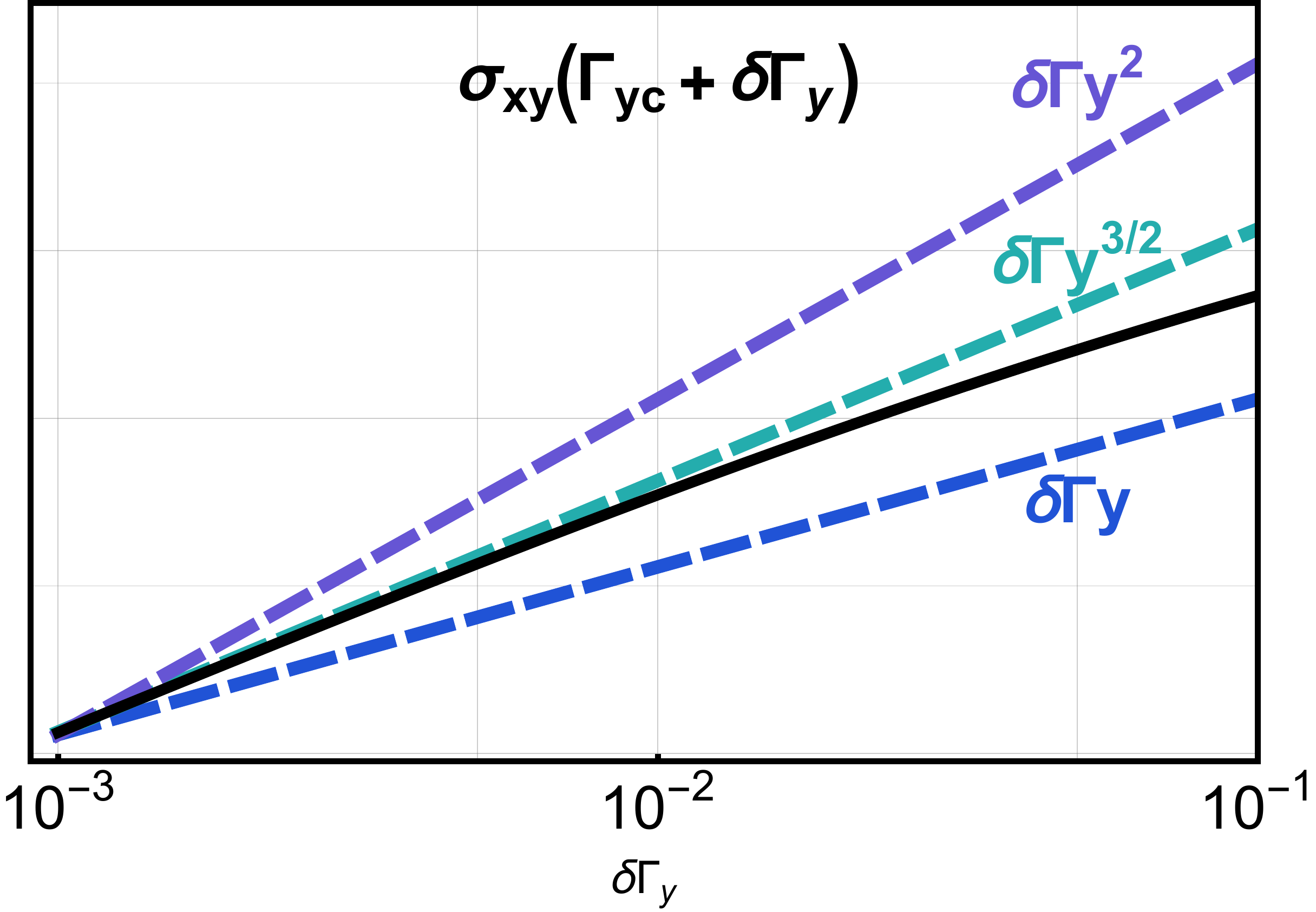} \\
\includegraphics[width=4cm]{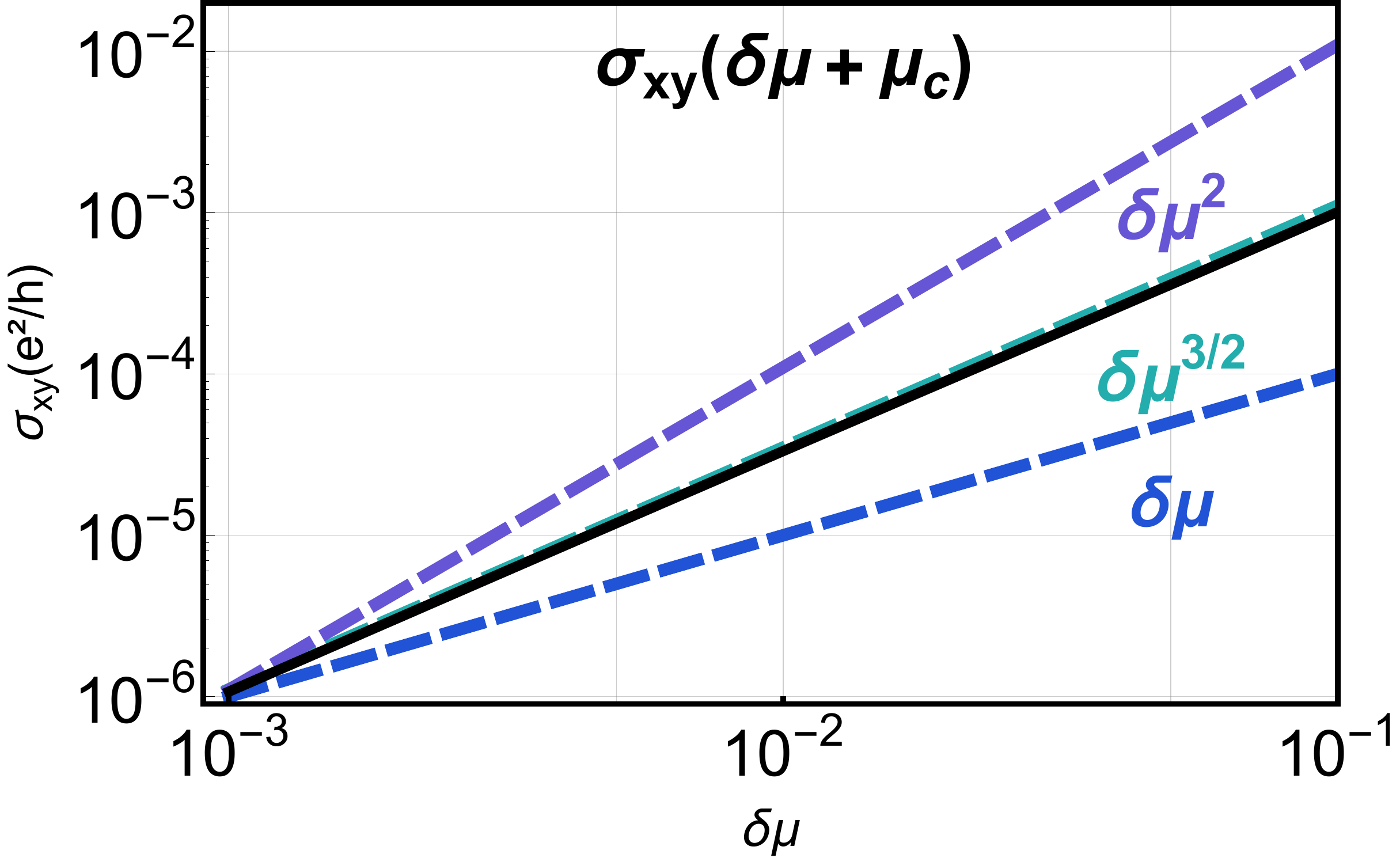} &
\end{tabular}
 \caption{The universal scaling of the Hall conductance contribution $\sigma^{(3/2)}_{xy}$ as a function of the system parameters $m$, $\Gamma_y$ and $\mu$. Here, $m_c$, $\Gamma_{yc}$ and $\mu_c$ denote, respectively, the critical values of the mass, the non-Hermiticity, and the chemical potential for which $\epsilon=1$. The exact function $\sigma^{(3/2)}_{xy}$ (solid line) is then compared to different power laws (dashed lines), showing that it scales indeed with an exponent $3/2$ as a function of these system parameters. The remaining parameters are $\mu=0.5$, $\Gamma_y=1$ and $m=1$.}
 \label{fig:onset}
\end{figure}

To explicitly evaluate Eq.~\eqref{eq:contrib}, we proceed by performing the integration over $k_x$, which yields
\begin{equation}\label{eq:epConduc}
\sigma^{(3/2)}_{xy} = \frac{e^2}{h} \frac{m}{\pi}\int_{\Gamma_0}^{\epsilon\Gamma_0} d\omega\ \Re\frac{\Theta(\epsilon-1) k_0(\omega)}{\left(\Gamma_y+i\omega\mu/\Gamma_y\right)\left[m^2+(\omega+i\mu)^2\right]}
\end{equation}
with $k_0(\omega)=\sqrt{(\Gamma_y^2+\mu^2-m^2)-\mu^2\omega^2/\Gamma_y^2-\omega^2}$. For $\mu=0$, this integral can be performed exactly, and we find near the onset (i.e.~for $\epsilon\approx 1$) that
\begin{equation}
\sigma^{(3/2)}_{xy}(\mu=0)\approx \frac{2e^2\sqrt{2} m }{3\pi h}\frac{\Gamma_y^2-m^2}{\Gamma_y^3} \Theta(\epsilon-1)(\epsilon -1)^{3/2}.
\end{equation}
The contribution $\sigma^{(3/2)}_{xy}$ to the Hall conductance thus scales with a power of $3/2$ with the distance to its onset given by $\epsilon$. This non-analytic behavior of the conductance is, however, by no means restricted to $\mu=0$. While we did not obtain a general closed result for the integral for $\mu>0$, analytic progress can be made for $\epsilon \approx 1$. We can then approximate $\omega \approx \Gamma_0$ and neglect terms containing $\omega-\Gamma_0$ in the denominator. The conductance can then be approximated as
\begin{align}
\sigma^{(3/2)}_{xy} &\approx \frac{e^2}{h} \frac{m}{\pi} \int_{\Gamma_0}^{\epsilon\Gamma_0}d\omega   \text{Re} \frac{\Theta(\epsilon -1)\sqrt{\epsilon^2\Gamma_0^2  -\omega^2}\sqrt{\mu^2/\Gamma_y^2+1 } }{\left(\Gamma_y+i\mu\Gamma_0/\Gamma_y\right)\left[m^2+(\Gamma_0+i\mu)^2\right]}  \notag \\
&\propto \Theta(\epsilon -1)(\epsilon-1)^{3/2}.
\end{align}
This shows that the Hall conductance generically contains a contribution that is non-analytic in the onset parameter $\epsilon$. We confirmed  this non-analytic onset scaling of $\sigma^{(3/2)}_{xy}$ by evaluating Eq.~\eqref{eq:epConduc} numerically upon variation of either the mass, the chemical potential or the non-Hermiticity $\Gamma_y$. This universal scaling is depicted in Fig.~\ref{fig:onset}.

\begin{figure}[t]
\centering
\includegraphics[width=\columnwidth]{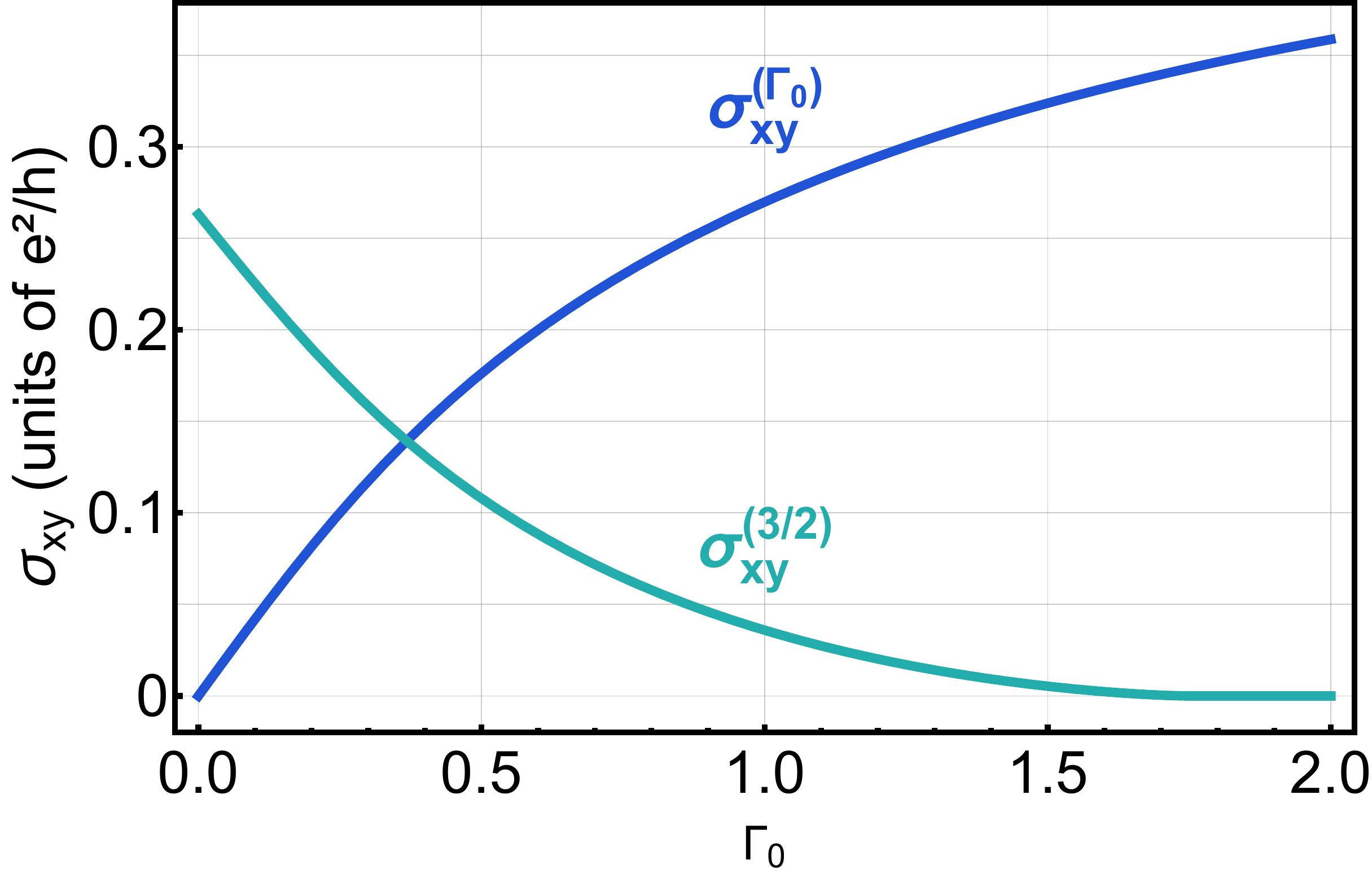}
 \caption{Contributions of Hall conductance
  as a function of the global loss term $\Gamma_0$ in units of $e^2/h$, in the regime where the system experiences both loss and gain ($\Gamma_y=2$), at finite chemical potential ($\mu=0.5$). The mass is taken to be unity. $\sigma_{xy}^{(\Gamma_0)}=1/2-|\tilde{\sigma}^{(sea)}_{xy}|$ is the Fermi sea contribution controlled by $\Gamma_0$ alone whereas $ \sigma^{(\rm 3/2)}_{xy}$ is given in the main text.}
 \label{fig:deviation}
\end{figure}

An important practical condition for a successful detection of $\sigma^{(3/2)}_{xy}$ is that it is not overwhelmed by the residual non-universal Hall conductance. In particular, the total Hall conductance is dominated by its non-universal contributions if the constant loss term $\Gamma_0$ is not much smaller than $\Gamma_y$, see Fig.~\ref{fig:deviation}. The most promising setup for observing the non-analytic contribution to the Hall conductance is thus vanishing chemical potential and compensated gains and losses (i.e.~$\mu=\Gamma_0=0$).

The non-analytic contribution to the Hall conductance identified in this work is fundamentally different from the non-universality of the Hall response of effective non-Hermitian Chern insulators resulting from the coupling to an environment on the level of self-energies \cite{Chen2018,Philip2018,Hirsbrunner2019,Wang2019}. Such kinds of effective non-Hermitian Hamiltonians have to respect the causality condition $\Gamma_0>|\Gamma_{x,y,z}|$. Physically, this condition encodes that self-energies are an effective description of a microscopically Hermitian system in which low-energy quasiparticles have a finite lifetime due to scattering into other states. As heralded by Eq.~\eqref{eq:condition}, the universal scaling identified in this work is in fact absent in such an effectively ``lossy'' non-Hermitian system. We thus conclude that non-Hermitian Chern insulators with $\Gamma_0>|\Gamma_{x,y,z}|$, in particular those deriving from complex self-energies, exhibit a  fundamentally unquantized and non-universal Hall response. In open systems, however, true loss and gain are a physical reality. This is already apparent on the level of Lindblad Master equations, for which a gain term of the form $i\Gamma_0\,\sigma_0+i\Gamma_y\,\sigma_y$ results from the combination of Lindblad operators $L_1 = \sqrt{(\Gamma_0+\Gamma_y)}\,(c_\uparrow-i\,c_\downarrow)$ and $L_2 = \sqrt{(-\Gamma_0+\Gamma_y)}\,(c_\uparrow^\dagger-i\,c_\downarrow^\dagger)$. The biorthogonal linear response theory we set up in the present work generalizes the discussion of non-Hermitian Hamiltonians from spectral properties to linear response functions.

In summary, we have analyzed a two-dimensional Chern insulator coupled to an environment that induces loss and gain of quasiparticles. Starting from the corresponding non-Hermitian Hamiltonian, we propose a generalized biorthogonal linear-response theory for non-Hermitian Hamiltonians, and show that this scheme can be connected to a Matsubara-like formalism via analytic continuation. In Chern insulators with loss, our scheme recovers results derived in the familiar language of complex self-energies stating that the  Hall conductance of a lossy Chern insulator is non-universal and not quantized. In addition, we also show that the Hall conductance is unaffected by the presence of exceptional points. Qualitatively new effects only emerge due to suitable combinations of loss and gain, which generally lead to the emergence of a new contribution to the Hall conductance. We demonstrated that near its onset, this new contribution is non-analytic and scales with a universal power $3/2$ as a function of all system parameters. This new form of non-analytic behavior in open topological quantum systems is independent of the singularities found near exceptional points in the complex spectrum of the effective Hamiltonian. Our findings thereby open a new avenue for using transport as a probe for non-Hermitian topological phases.

\begin{acknowledgments}
The authors acknowledge helpful discussions with J.~C.~Budich. SG and TLS acknowledge support by the National Research Fund, Luxembourg under grants ATTRACT 7556175 and PRIDE/15/10935404.
TM acknowledges financial support by the Deut\-sche For\-schungs\-ge\-mein\-schaft via the Emmy Noether Programme ME4844/1-1 (project id 327807255), the Collaborative Research Center SFB 1143 (project id 247310070), and the Cluster of Excellence on Complexity and Topology in Quantum Matter \textit{ct.qmat} (EXC 2147, project id 390858490).
\end{acknowledgments}

\bibliography{references}

\begin{thebibliography}{47}
\expandafter\ifx\csname natexlab\endcsname\relax\def\natexlab#1{#1}\fi
\expandafter\ifx\csname bibnamefont\endcsname\relax
  \def\bibnamefont#1{#1}\fi
\expandafter\ifx\csname bibfnamefont\endcsname\relax
  \def\bibfnamefont#1{#1}\fi
\expandafter\ifx\csname citenamefont\endcsname\relax
  \def\citenamefont#1{#1}\fi
\expandafter\ifx\csname url\endcsname\relax
  \def\url#1{\texttt{#1}}\fi
\expandafter\ifx\csname urlprefix\endcsname\relax\def\urlprefix{URL }\fi
\providecommand{\bibinfo}[2]{#2}
\providecommand{\eprint}[2][]{\url{#2}}

\bibitem[{\citenamefont{Kosterlitz and Thouless}(1973)}]{Kosterlitz1973}
\bibinfo{author}{\bibfnamefont{J.~M.} \bibnamefont{Kosterlitz}}
  \bibnamefont{and} \bibinfo{author}{\bibfnamefont{D.~J.}
  \bibnamefont{Thouless}}, \bibinfo{journal}{J. Phys. C: Solid State Phys.}
  \textbf{\bibinfo{volume}{6}}, \bibinfo{pages}{1181} (\bibinfo{year}{1973}).

\bibitem[{\citenamefont{Wen}(2017)}]{Wen2017}
\bibinfo{author}{\bibfnamefont{X.-G.} \bibnamefont{Wen}},
  \bibinfo{journal}{Rev. Mod. Phys.} \textbf{\bibinfo{volume}{89}},
  \bibinfo{pages}{041004} (\bibinfo{year}{2017}).

\bibitem[{\citenamefont{Chiu et~al.}(2016)\citenamefont{Chiu, Teo, Schnyder,
  and Ryu}}]{Chiu2016}
\bibinfo{author}{\bibfnamefont{C.-K.} \bibnamefont{Chiu}},
  \bibinfo{author}{\bibfnamefont{J.~C.} \bibnamefont{Teo}},
  \bibinfo{author}{\bibfnamefont{A.~P.} \bibnamefont{Schnyder}},
  \bibnamefont{and} \bibinfo{author}{\bibfnamefont{S.}~\bibnamefont{Ryu}},
  \bibinfo{journal}{Rev. Mod. Phys.} \textbf{\bibinfo{volume}{88}},
  \bibinfo{pages}{035005} (\bibinfo{year}{2016}).

\bibitem[{\citenamefont{Qi and Zhang}(2011)}]{Qi2011}
\bibinfo{author}{\bibfnamefont{X.-L.} \bibnamefont{Qi}} \bibnamefont{and}
  \bibinfo{author}{\bibfnamefont{S.-C.} \bibnamefont{Zhang}},
  \bibinfo{journal}{Rev. Mod. Phys.} \textbf{\bibinfo{volume}{83}},
  \bibinfo{pages}{1057} (\bibinfo{year}{2011}).

\bibitem[{\citenamefont{Kitaev et~al.}(2009)\citenamefont{Kitaev, Lebedev, and
  Feigel'man}}]{Kitaev2009}
\bibinfo{author}{\bibfnamefont{A.}~\bibnamefont{Kitaev}},
  \bibinfo{author}{\bibfnamefont{V.}~\bibnamefont{Lebedev}}, \bibnamefont{and}
  \bibinfo{author}{\bibfnamefont{M.}~\bibnamefont{Feigel'man}},
  \bibinfo{journal}{AIP Conf. Proc.} \textbf{\bibinfo{volume}{1134}},
  \bibinfo{pages}{22} (\bibinfo{year}{2009}).

\bibitem[{\citenamefont{Thouless et~al.}(1982)\citenamefont{Thouless, Kohmoto,
  Nightingale, and den Nijs}}]{Thouless1982}
\bibinfo{author}{\bibfnamefont{D.~J.} \bibnamefont{Thouless}},
  \bibinfo{author}{\bibfnamefont{M.}~\bibnamefont{Kohmoto}},
  \bibinfo{author}{\bibfnamefont{M.~P.} \bibnamefont{Nightingale}},
  \bibnamefont{and} \bibinfo{author}{\bibfnamefont{M.}~\bibnamefont{den Nijs}},
  \bibinfo{journal}{Phys. Rev. Lett.} \textbf{\bibinfo{volume}{49}},
  \bibinfo{pages}{405} (\bibinfo{year}{1982}).

\bibitem[{\citenamefont{Haldane}(1988)}]{Haldane1988}
\bibinfo{author}{\bibfnamefont{F.~D.~M.} \bibnamefont{Haldane}},
  \bibinfo{journal}{Phys. Rev. Lett.} \textbf{\bibinfo{volume}{61}},
  \bibinfo{pages}{2015} (\bibinfo{year}{1988}).

\bibitem[{\citenamefont{Liu et~al.}(2016)\citenamefont{Liu, Zhang, and
  Qi}}]{Liu2016}
\bibinfo{author}{\bibfnamefont{C.-X.} \bibnamefont{Liu}},
  \bibinfo{author}{\bibfnamefont{S.-C.} \bibnamefont{Zhang}}, \bibnamefont{and}
  \bibinfo{author}{\bibfnamefont{X.-L.} \bibnamefont{Qi}},
  \bibinfo{journal}{Annu. Rev. Condens. Matter Phys}
  \textbf{\bibinfo{volume}{7}}, \bibinfo{pages}{301} (\bibinfo{year}{2016}).

\bibitem[{\citenamefont{Gulden et~al.}(2016)\citenamefont{Gulden, Janas, Wang,
  and Kamenev}}]{Gulden2016}
\bibinfo{author}{\bibfnamefont{T.}~\bibnamefont{Gulden}},
  \bibinfo{author}{\bibfnamefont{M.}~\bibnamefont{Janas}},
  \bibinfo{author}{\bibfnamefont{Y.}~\bibnamefont{Wang}}, \bibnamefont{and}
  \bibinfo{author}{\bibfnamefont{A.}~\bibnamefont{Kamenev}},
  \bibinfo{journal}{Phys. Rev. Lett.} \textbf{\bibinfo{volume}{116}},
  \bibinfo{pages}{026402} (\bibinfo{year}{2016}).

\bibitem[{\citenamefont{Zhou et~al.}(2008)\citenamefont{Zhou, Lu, Chu, Shen,
  and Niu}}]{Zhou2008}
\bibinfo{author}{\bibfnamefont{B.}~\bibnamefont{Zhou}},
  \bibinfo{author}{\bibfnamefont{H.-Z.} \bibnamefont{Lu}},
  \bibinfo{author}{\bibfnamefont{R.-L.} \bibnamefont{Chu}},
  \bibinfo{author}{\bibfnamefont{S.-Q.} \bibnamefont{Shen}}, \bibnamefont{and}
  \bibinfo{author}{\bibfnamefont{Q.}~\bibnamefont{Niu}},
  \bibinfo{journal}{Phys. Rev. Lett.} \textbf{\bibinfo{volume}{101}},
  \bibinfo{pages}{246807} (\bibinfo{year}{2008}).

\bibitem[{\citenamefont{Albrecht et~al.}(2016)\citenamefont{Albrecht,
  Higginbotham, Madsen, Kuemmeth, Jespersen, Nyg{\aa}rd, Krogstrup, and
  Marcus}}]{Albrecht2016}
\bibinfo{author}{\bibfnamefont{S.~M.} \bibnamefont{Albrecht}},
  \bibinfo{author}{\bibfnamefont{A.~P.} \bibnamefont{Higginbotham}},
  \bibinfo{author}{\bibfnamefont{M.}~\bibnamefont{Madsen}},
  \bibinfo{author}{\bibfnamefont{F.}~\bibnamefont{Kuemmeth}},
  \bibinfo{author}{\bibfnamefont{T.~S.} \bibnamefont{Jespersen}},
  \bibinfo{author}{\bibfnamefont{J.}~\bibnamefont{Nyg{\aa}rd}},
  \bibinfo{author}{\bibfnamefont{P.}~\bibnamefont{Krogstrup}},
  \bibnamefont{and} \bibinfo{author}{\bibfnamefont{C.~M.}
  \bibnamefont{Marcus}}, \bibinfo{journal}{Nature}
  \textbf{\bibinfo{volume}{531}}, \bibinfo{pages}{206} (\bibinfo{year}{2016}).

\bibitem[{\citenamefont{Giusteri et~al.}(2015)\citenamefont{Giusteri,
  Mattiotti, and Celardo}}]{Giusteri2015}
\bibinfo{author}{\bibfnamefont{G.~G.} \bibnamefont{Giusteri}},
  \bibinfo{author}{\bibfnamefont{F.}~\bibnamefont{Mattiotti}},
  \bibnamefont{and} \bibinfo{author}{\bibfnamefont{G.~L.}
  \bibnamefont{Celardo}}, \bibinfo{journal}{Phys. Rev. B}
  \textbf{\bibinfo{volume}{91}}, \bibinfo{pages}{094301}
  (\bibinfo{year}{2015}).

\bibitem[{\citenamefont{Bergholtz et~al.}(2019)\citenamefont{Bergholtz, Budich,
  and Kunst}}]{Bergholtz2019}
\bibinfo{author}{\bibfnamefont{E.~J.} \bibnamefont{Bergholtz}},
  \bibinfo{author}{\bibfnamefont{J.~C.} \bibnamefont{Budich}},
  \bibnamefont{and} \bibinfo{author}{\bibfnamefont{F.~K.} \bibnamefont{Kunst}},
  \bibinfo{journal}{arXiv:1912.10048v2}  (\bibinfo{year}{2019}),
  \eprint{1912.10048v2}.

\bibitem[{\citenamefont{Torres}(2019)}]{Torres2019}
\bibinfo{author}{\bibfnamefont{L.~E. F.~F.} \bibnamefont{Torres}},
  \bibinfo{journal}{J. Phys. Mater.} \textbf{\bibinfo{volume}{3}},
  \bibinfo{pages}{014002} (\bibinfo{year}{2019}).

\bibitem[{sup()}]{supplement}
\bibinfo{note}{See Supplemental Material.}

\bibitem[{\citenamefont{Philip et~al.}(2018)\citenamefont{Philip, Hirsbrunner,
  and Gilbert}}]{Philip2018}
\bibinfo{author}{\bibfnamefont{T.~M.} \bibnamefont{Philip}},
  \bibinfo{author}{\bibfnamefont{M.~R.} \bibnamefont{Hirsbrunner}},
  \bibnamefont{and} \bibinfo{author}{\bibfnamefont{M.~J.}
  \bibnamefont{Gilbert}}, \bibinfo{journal}{Phys. Rev. B}
  \textbf{\bibinfo{volume}{98}}, \bibinfo{pages}{155430}
  (\bibinfo{year}{2018}).

\bibitem[{\citenamefont{Hirsbrunner et~al.}(2019)\citenamefont{Hirsbrunner,
  Philip, and Gilbert}}]{Hirsbrunner2019}
\bibinfo{author}{\bibfnamefont{M.~R.} \bibnamefont{Hirsbrunner}},
  \bibinfo{author}{\bibfnamefont{T.~M.} \bibnamefont{Philip}},
  \bibnamefont{and} \bibinfo{author}{\bibfnamefont{M.~J.}
  \bibnamefont{Gilbert}}, \bibinfo{journal}{Phys. Rev. B}
  \textbf{\bibinfo{volume}{100}}, \bibinfo{pages}{081104}
  (\bibinfo{year}{2019}).

\bibitem[{\citenamefont{Berry}(2004)}]{Berry2004}
\bibinfo{author}{\bibfnamefont{M.}~\bibnamefont{Berry}},
  \bibinfo{journal}{Czech. J. Phys.} \textbf{\bibinfo{volume}{54}},
  \bibinfo{pages}{1039} (\bibinfo{year}{2004}).

\bibitem[{\citenamefont{Rotter}(2009)}]{Rotter2009}
\bibinfo{author}{\bibfnamefont{I.}~\bibnamefont{Rotter}}, \bibinfo{journal}{J.
  Phys. A: Math. Theor.} \textbf{\bibinfo{volume}{42}}, \bibinfo{pages}{153001}
  (\bibinfo{year}{2009}).

\bibitem[{\citenamefont{Shen et~al.}(2018)\citenamefont{Shen, Zhen, and
  Fu}}]{Shen2018}
\bibinfo{author}{\bibfnamefont{H.}~\bibnamefont{Shen}},
  \bibinfo{author}{\bibfnamefont{B.}~\bibnamefont{Zhen}}, \bibnamefont{and}
  \bibinfo{author}{\bibfnamefont{L.}~\bibnamefont{Fu}}, \bibinfo{journal}{Phys.
  Rev. Lett.} \textbf{\bibinfo{volume}{120}}, \bibinfo{pages}{146402}
  (\bibinfo{year}{2018}).

\bibitem[{\citenamefont{Papaj et~al.}(2019)\citenamefont{Papaj, Isobe, and
  Fu}}]{Papaj2019}
\bibinfo{author}{\bibfnamefont{M.}~\bibnamefont{Papaj}},
  \bibinfo{author}{\bibfnamefont{H.}~\bibnamefont{Isobe}}, \bibnamefont{and}
  \bibinfo{author}{\bibfnamefont{L.}~\bibnamefont{Fu}}, \bibinfo{journal}{Phys.
  Rev. B} \textbf{\bibinfo{volume}{99}}, \bibinfo{pages}{201107}
  (\bibinfo{year}{2019}).

\bibitem[{\citenamefont{Diehl et~al.}(2011)\citenamefont{Diehl, Rico, Baranov,
  and Zoller}}]{Diehl2011}
\bibinfo{author}{\bibfnamefont{S.}~\bibnamefont{Diehl}},
  \bibinfo{author}{\bibfnamefont{E.}~\bibnamefont{Rico}},
  \bibinfo{author}{\bibfnamefont{M.~A.} \bibnamefont{Baranov}},
  \bibnamefont{and} \bibinfo{author}{\bibfnamefont{P.}~\bibnamefont{Zoller}},
  \bibinfo{journal}{Nat. Phys.} \textbf{\bibinfo{volume}{7}},
  \bibinfo{pages}{971} (\bibinfo{year}{2011}).

\bibitem[{\citenamefont{El-Ganainy et~al.}(2018)\citenamefont{El-Ganainy,
  Makris, Khajavikhan, Musslimani, Rotter, and
  Christodoulides}}]{El-Ganainy2018}
\bibinfo{author}{\bibfnamefont{R.}~\bibnamefont{El-Ganainy}},
  \bibinfo{author}{\bibfnamefont{K.~G.} \bibnamefont{Makris}},
  \bibinfo{author}{\bibfnamefont{M.}~\bibnamefont{Khajavikhan}},
  \bibinfo{author}{\bibfnamefont{Z.~H.} \bibnamefont{Musslimani}},
  \bibinfo{author}{\bibfnamefont{S.}~\bibnamefont{Rotter}}, \bibnamefont{and}
  \bibinfo{author}{\bibfnamefont{D.~N.} \bibnamefont{Christodoulides}},
  \bibinfo{journal}{Nat. Phys.} \textbf{\bibinfo{volume}{14}},
  \bibinfo{pages}{11} (\bibinfo{year}{2018}).

\bibitem[{\citenamefont{Song et~al.}(2019)\citenamefont{Song, Yao, and
  Wang}}]{Song2019}
\bibinfo{author}{\bibfnamefont{F.}~\bibnamefont{Song}},
  \bibinfo{author}{\bibfnamefont{S.}~\bibnamefont{Yao}}, \bibnamefont{and}
  \bibinfo{author}{\bibfnamefont{Z.}~\bibnamefont{Wang}},
  \bibinfo{journal}{Phys. Rev. Lett.} \textbf{\bibinfo{volume}{123}},
  \bibinfo{pages}{170401} (\bibinfo{year}{2019}).

\bibitem[{\citenamefont{Naghiloo et~al.}(2019)\citenamefont{Naghiloo, Abbasi,
  Joglekar, and Murch}}]{Naghiloo2019}
\bibinfo{author}{\bibfnamefont{M.}~\bibnamefont{Naghiloo}},
  \bibinfo{author}{\bibfnamefont{M.}~\bibnamefont{Abbasi}},
  \bibinfo{author}{\bibfnamefont{Y.~N.} \bibnamefont{Joglekar}},
  \bibnamefont{and} \bibinfo{author}{\bibfnamefont{K.~W.} \bibnamefont{Murch}},
  \bibinfo{journal}{Nat. Phys.} \textbf{\bibinfo{volume}{15}},
  \bibinfo{pages}{1232} (\bibinfo{year}{2019}).

\bibitem[{\citenamefont{Minganti et~al.}(2019)\citenamefont{Minganti,
  Miranowicz, Chhajlany, and Nori}}]{Minganti2019}
\bibinfo{author}{\bibfnamefont{F.}~\bibnamefont{Minganti}},
  \bibinfo{author}{\bibfnamefont{A.}~\bibnamefont{Miranowicz}},
  \bibinfo{author}{\bibfnamefont{R.~W.} \bibnamefont{Chhajlany}},
  \bibnamefont{and} \bibinfo{author}{\bibfnamefont{F.}~\bibnamefont{Nori}},
  \bibinfo{journal}{Phys. Rev. A} \textbf{\bibinfo{volume}{100}},
  \bibinfo{pages}{062131} (\bibinfo{year}{2019}).

\bibitem[{\citenamefont{Tripathi and Vinokur}(2020)}]{Tripathi2020}
\bibinfo{author}{\bibfnamefont{V.}~\bibnamefont{Tripathi}} \bibnamefont{and}
  \bibinfo{author}{\bibfnamefont{V.~M.} \bibnamefont{Vinokur}},
  \bibinfo{journal}{Sci. Rep.} \textbf{\bibinfo{volume}{10}},
  \bibinfo{pages}{7304} (\bibinfo{year}{2020}).

\bibitem[{\citenamefont{Brody}(2013)}]{Brody2013}
\bibinfo{author}{\bibfnamefont{D.~C.} \bibnamefont{Brody}},
  \bibinfo{journal}{J. Phys. A: Math. Theor.} \textbf{\bibinfo{volume}{47}},
  \bibinfo{pages}{035305} (\bibinfo{year}{2013}).

\bibitem[{\citenamefont{Herviou et~al.}(2019)\citenamefont{Herviou, Regnault,
  and Bardarson}}]{Herviou2019}
\bibinfo{author}{\bibfnamefont{L.}~\bibnamefont{Herviou}},
  \bibinfo{author}{\bibfnamefont{N.}~\bibnamefont{Regnault}}, \bibnamefont{and}
  \bibinfo{author}{\bibfnamefont{J.~H.} \bibnamefont{Bardarson}},
  \bibinfo{journal}{SciPost Phys.} \textbf{\bibinfo{volume}{7}},
  \bibinfo{pages}{69} (\bibinfo{year}{2019}).

\bibitem[{\citenamefont{Regensburger et~al.}(2012)\citenamefont{Regensburger,
  Bersch, Miri, Onishchukov, Christodoulides, and Peschel}}]{Regensburger2012}
\bibinfo{author}{\bibfnamefont{A.}~\bibnamefont{Regensburger}},
  \bibinfo{author}{\bibfnamefont{C.}~\bibnamefont{Bersch}},
  \bibinfo{author}{\bibfnamefont{M.-A.} \bibnamefont{Miri}},
  \bibinfo{author}{\bibfnamefont{G.}~\bibnamefont{Onishchukov}},
  \bibinfo{author}{\bibfnamefont{D.~N.} \bibnamefont{Christodoulides}},
  \bibnamefont{and} \bibinfo{author}{\bibfnamefont{U.}~\bibnamefont{Peschel}},
  \bibinfo{journal}{Nature} \textbf{\bibinfo{volume}{488}},
  \bibinfo{pages}{167} (\bibinfo{year}{2012}).

\bibitem[{\citenamefont{Zhen et~al.}(2015)\citenamefont{Zhen, Hsu, Igarashi,
  Lu, Kaminer, Pick, Chua, Joannopoulos, and Solja{\v{c}}i{\'{c}}}}]{Zhen2015}
\bibinfo{author}{\bibfnamefont{B.}~\bibnamefont{Zhen}},
  \bibinfo{author}{\bibfnamefont{C.~W.} \bibnamefont{Hsu}},
  \bibinfo{author}{\bibfnamefont{Y.}~\bibnamefont{Igarashi}},
  \bibinfo{author}{\bibfnamefont{L.}~\bibnamefont{Lu}},
  \bibinfo{author}{\bibfnamefont{I.}~\bibnamefont{Kaminer}},
  \bibinfo{author}{\bibfnamefont{A.}~\bibnamefont{Pick}},
  \bibinfo{author}{\bibfnamefont{S.-L.} \bibnamefont{Chua}},
  \bibinfo{author}{\bibfnamefont{J.~D.} \bibnamefont{Joannopoulos}},
  \bibnamefont{and}
  \bibinfo{author}{\bibfnamefont{M.}~\bibnamefont{Solja{\v{c}}i{\'{c}}}},
  \bibinfo{journal}{Nature} \textbf{\bibinfo{volume}{525}},
  \bibinfo{pages}{354} (\bibinfo{year}{2015}).

\bibitem[{\citenamefont{Hahn et~al.}(2016)\citenamefont{Hahn, Choi, Yoon, Song,
  Oh, and Berini}}]{Hahn2016}
\bibinfo{author}{\bibfnamefont{C.}~\bibnamefont{Hahn}},
  \bibinfo{author}{\bibfnamefont{Y.}~\bibnamefont{Choi}},
  \bibinfo{author}{\bibfnamefont{J.~W.} \bibnamefont{Yoon}},
  \bibinfo{author}{\bibfnamefont{S.~H.} \bibnamefont{Song}},
  \bibinfo{author}{\bibfnamefont{C.~H.} \bibnamefont{Oh}}, \bibnamefont{and}
  \bibinfo{author}{\bibfnamefont{P.}~\bibnamefont{Berini}},
  \bibinfo{journal}{Nat. Commun.} \textbf{\bibinfo{volume}{7}},
  \bibinfo{pages}{12201} (\bibinfo{year}{2016}).

\bibitem[{\citenamefont{Özdemir et~al.}(2019)\citenamefont{Özdemir, Rotter,
  Nori, and Yang}}]{Oezdemir2019}
\bibinfo{author}{\bibfnamefont{{\c{S}}.~K.} \bibnamefont{Özdemir}},
  \bibinfo{author}{\bibfnamefont{S.}~\bibnamefont{Rotter}},
  \bibinfo{author}{\bibfnamefont{F.}~\bibnamefont{Nori}}, \bibnamefont{and}
  \bibinfo{author}{\bibfnamefont{L.}~\bibnamefont{Yang}},
  \bibinfo{journal}{Nat. Mater.} \textbf{\bibinfo{volume}{18}},
  \bibinfo{pages}{783} (\bibinfo{year}{2019}).

\bibitem[{\citenamefont{Miri and Al{\`{u}}}(2019)}]{Miri2019}
\bibinfo{author}{\bibfnamefont{M.-A.} \bibnamefont{Miri}} \bibnamefont{and}
  \bibinfo{author}{\bibfnamefont{A.}~\bibnamefont{Al{\`{u}}}},
  \bibinfo{journal}{Science} \textbf{\bibinfo{volume}{363}},
  \bibinfo{pages}{7709} (\bibinfo{year}{2019}).

\bibitem[{\citenamefont{Huang et~al.}(2017)\citenamefont{Huang, Shen, Min, Fan,
  and Veronis}}]{Huang2017}
\bibinfo{author}{\bibfnamefont{Y.}~\bibnamefont{Huang}},
  \bibinfo{author}{\bibfnamefont{Y.}~\bibnamefont{Shen}},
  \bibinfo{author}{\bibfnamefont{C.}~\bibnamefont{Min}},
  \bibinfo{author}{\bibfnamefont{S.}~\bibnamefont{Fan}}, \bibnamefont{and}
  \bibinfo{author}{\bibfnamefont{G.}~\bibnamefont{Veronis}},
  \bibinfo{journal}{Nanophotonics} \textbf{\bibinfo{volume}{6}},
  \bibinfo{pages}{977} (\bibinfo{year}{2017}).

\bibitem[{\citenamefont{Guo et~al.}(2009)\citenamefont{Guo, Salamo, Duchesne,
  Morandotti, Volatier-Ravat, Aimez, Siviloglou, and
  Christodoulides}}]{Guo2009}
\bibinfo{author}{\bibfnamefont{A.}~\bibnamefont{Guo}},
  \bibinfo{author}{\bibfnamefont{G.~J.} \bibnamefont{Salamo}},
  \bibinfo{author}{\bibfnamefont{D.}~\bibnamefont{Duchesne}},
  \bibinfo{author}{\bibfnamefont{R.}~\bibnamefont{Morandotti}},
  \bibinfo{author}{\bibfnamefont{M.}~\bibnamefont{Volatier-Ravat}},
  \bibinfo{author}{\bibfnamefont{V.}~\bibnamefont{Aimez}},
  \bibinfo{author}{\bibfnamefont{G.~A.} \bibnamefont{Siviloglou}},
  \bibnamefont{and} \bibinfo{author}{\bibfnamefont{D.~N.}
  \bibnamefont{Christodoulides}}, \bibinfo{journal}{Phys. Rev. Lett.}
  \textbf{\bibinfo{volume}{103}}, \bibinfo{pages}{093902}
  (\bibinfo{year}{2009}).

\bibitem[{\citenamefont{Hodaei et~al.}(2017)\citenamefont{Hodaei, Hassan,
  Wittek, Garcia-Gracia, El-Ganainy, Christodoulides, and
  Khajavikhan}}]{Hodaei2017}
\bibinfo{author}{\bibfnamefont{H.}~\bibnamefont{Hodaei}},
  \bibinfo{author}{\bibfnamefont{A.~U.} \bibnamefont{Hassan}},
  \bibinfo{author}{\bibfnamefont{S.}~\bibnamefont{Wittek}},
  \bibinfo{author}{\bibfnamefont{H.}~\bibnamefont{Garcia-Gracia}},
  \bibinfo{author}{\bibfnamefont{R.}~\bibnamefont{El-Ganainy}},
  \bibinfo{author}{\bibfnamefont{D.~N.} \bibnamefont{Christodoulides}},
  \bibnamefont{and}
  \bibinfo{author}{\bibfnamefont{M.}~\bibnamefont{Khajavikhan}},
  \bibinfo{journal}{Nature} \textbf{\bibinfo{volume}{548}},
  \bibinfo{pages}{187} (\bibinfo{year}{2017}).

\bibitem[{\citenamefont{Ishikawa and Matsuyama}(1986)}]{Ishikawa1986}
\bibinfo{author}{\bibfnamefont{K.}~\bibnamefont{Ishikawa}} \bibnamefont{and}
  \bibinfo{author}{\bibfnamefont{T.}~\bibnamefont{Matsuyama}},
  \bibinfo{journal}{Z. Phys. C} \textbf{\bibinfo{volume}{33}},
  \bibinfo{pages}{41} (\bibinfo{year}{1986}).

\bibitem[{\citenamefont{Ishikawa and Matsuyama}(1987)}]{Ishikawa1987}
\bibinfo{author}{\bibfnamefont{K.}~\bibnamefont{Ishikawa}} \bibnamefont{and}
  \bibinfo{author}{\bibfnamefont{T.}~\bibnamefont{Matsuyama}},
  \bibinfo{journal}{Nucl. Phys. B} \textbf{\bibinfo{volume}{280}},
  \bibinfo{pages}{523} (\bibinfo{year}{1987}).

\bibitem[{\citenamefont{Imai et~al.}(1990)\citenamefont{Imai, Ishikawa,
  Matsuyama, and Tanaka}}]{Imai1990}
\bibinfo{author}{\bibfnamefont{N.}~\bibnamefont{Imai}},
  \bibinfo{author}{\bibfnamefont{K.}~\bibnamefont{Ishikawa}},
  \bibinfo{author}{\bibfnamefont{T.}~\bibnamefont{Matsuyama}},
  \bibnamefont{and} \bibinfo{author}{\bibfnamefont{I.}~\bibnamefont{Tanaka}},
  \bibinfo{journal}{Phys. Rev. B} \textbf{\bibinfo{volume}{42}},
  \bibinfo{pages}{10610} (\bibinfo{year}{1990}).

\bibitem[{\citenamefont{Wang et~al.}(2010)\citenamefont{Wang, Qi, and
  Zhang}}]{Wang2010}
\bibinfo{author}{\bibfnamefont{Z.}~\bibnamefont{Wang}},
  \bibinfo{author}{\bibfnamefont{X.-L.} \bibnamefont{Qi}}, \bibnamefont{and}
  \bibinfo{author}{\bibfnamefont{S.-C.} \bibnamefont{Zhang}},
  \bibinfo{journal}{Phys. Rev. Lett.} \textbf{\bibinfo{volume}{105}},
  \bibinfo{pages}{256803} (\bibinfo{year}{2010}).

\bibitem[{\citenamefont{Tsvelik}(2003)}]{Tsvelik2003}
\bibinfo{author}{\bibfnamefont{A.}~\bibnamefont{Tsvelik}},
  \emph{\bibinfo{title}{Quantum field theory in condensed matter physics}}
  (\bibinfo{publisher}{Cambridge University Press}, \bibinfo{address}{Cambridge
  New York}, \bibinfo{year}{2003}), ISBN \bibinfo{isbn}{9780511615832}.

\bibitem[{\citenamefont{Altland}(2010)}]{Altland2010}
\bibinfo{author}{\bibfnamefont{A.}~\bibnamefont{Altland}},
  \emph{\bibinfo{title}{Condensed matter field theory}}
  (\bibinfo{publisher}{Cambridge University Press}, \bibinfo{address}{Leiden},
  \bibinfo{year}{2010}), ISBN \bibinfo{isbn}{9780511789984}.

\bibitem[{\citenamefont{Belavin et~al.}(1975)\citenamefont{Belavin, Polyakov,
  Schwartz, and Tyupkin}}]{Belavin1975}
\bibinfo{author}{\bibfnamefont{A.}~\bibnamefont{Belavin}},
  \bibinfo{author}{\bibfnamefont{A.}~\bibnamefont{Polyakov}},
  \bibinfo{author}{\bibfnamefont{A.}~\bibnamefont{Schwartz}}, \bibnamefont{and}
  \bibinfo{author}{\bibfnamefont{Y.}~\bibnamefont{Tyupkin}},
  \bibinfo{journal}{Phys. Lett. B} \textbf{\bibinfo{volume}{59}},
  \bibinfo{pages}{85} (\bibinfo{year}{1975}).

\bibitem[{\citenamefont{Chern and Simons}(1974)}]{Chern1974}
\bibinfo{author}{\bibfnamefont{S.-S.} \bibnamefont{Chern}} \bibnamefont{and}
  \bibinfo{author}{\bibfnamefont{J.}~\bibnamefont{Simons}},
  \bibinfo{journal}{Ann. Math} \textbf{\bibinfo{volume}{99}},
  \bibinfo{pages}{48} (\bibinfo{year}{1974}).

\bibitem[{\citenamefont{Chen and Zhai}(2018)}]{Chen2018}
\bibinfo{author}{\bibfnamefont{Y.}~\bibnamefont{Chen}} \bibnamefont{and}
  \bibinfo{author}{\bibfnamefont{H.}~\bibnamefont{Zhai}},
  \bibinfo{journal}{Phys. Rev. B} \textbf{\bibinfo{volume}{98}},
  \bibinfo{pages}{245130} (\bibinfo{year}{2018}).

\bibitem[{\citenamefont{Wang and Wang}(2019)}]{Wang2019}
\bibinfo{author}{\bibfnamefont{C.}~\bibnamefont{Wang}} \bibnamefont{and}
  \bibinfo{author}{\bibfnamefont{X.~R.} \bibnamefont{Wang}},
  \bibinfo{journal}{arXiv:1901.06982}  (\bibinfo{year}{2019}),
  \eprint{1901.06982v1}.

\end{thebibliography}

\clearpage
\onecolumngrid
\appendix

\begin{center}
  \Large \bf Supplemental material
\end{center}

\section{Biorthogonal linear response theory}

Biorthogonal quantum mechanics has been reviewed in detail in Ref.~\cite{Brody2013}. For the derivation of the Matsubara technique and linear response theory, it is sufficient to take into account that (a) there exists a unique mapping between a given right eigenstate $\ket{\phi_n}$ and its corresponding left eigenstate, labelled $\ket{\tphi_n}$, (b) left and right eigenstates can be chosen as pairwise orthonormal, $\braket{\tphi_m|\phi_n} = \delta_{mn}$, and (c) these states together allow a biorthogonal completeness relation $\mathds{1} = \sum_n \ket{\phi_n} \bra{\tphi_n}$.

We define the correlation function of two operators $A$ and $B$ in Matsubara space as
\begin{align}
    C_{AB}^M(\tau) &= - \Tr' \left[ \rho T_\tau A(\tau) B(0) \right].
\end{align}
where $\rho = e^{-\beta H_{\rm nh}}/Z$ is the density operator of the system, with partition function $Z = \Tr'e^{-\beta H_{\rm nh}}$. Moreover, $T_\tau$ denotes the usual imaginary-time-ordering. Importantly, the time evolution of the operators is given by $A(\tau) = e^{H_{\rm nh} \tau} A e^{-H_{\rm nh} \tau}$. These definitions make it possible to recover most results of the conventional Matsubara technique. In particular, the biorthogonal completeness relation allows a Lehmann spectral representation resulting in
\begin{align}
    C_{AB}^M(i \omega_n) &= \int_0^\beta d \tau e^{i \omega_n \tau} C_{AB}^M(\tau) = \frac{1}{Z} \sum_{m,n} \bra{\tphi_m} A \ket{\phi_n} \bra{\tphi_n} B \ket{\phi_m} \frac{e^{-\beta\epsilon_m } - e^{-\beta\epsilon_n }}{i \omega_n + \epsilon_m - \epsilon_n}.
\end{align}

The aim of linear response theory is to calculate the expectation value of an operator $A$ to the first order in an external perturbation which couples to a system operator $B$. This corresponds to calculating the expectation value,
\begin{align}
    \braket{A}(t) = \Tr'[\rho(t) A],
\end{align}
where $\rho(t)$ is the time-evolved (biorthogonal) density matrix $\rho(t) = \sum_n e^{-\beta \epsilon_n} \ket{\phi_n(t)}\bra{\tphi_n(t)}$, where the time evolution of $\ket{\phi_n}$ and $\ket{\tphi_n}$ is given in the main text. Moreover, the external perturbation is switched on at time $t_0$ and leads to an additional term $V(t) = \theta(t-t_0) f(t) B$ in the Hamiltonian. Treating this term as a perturbation, we pass to the interaction picture as usually,
\begin{align}
    \braket{A}(t) &= \Tr'[\rho_0 A(t)] + \braket{\delta A}(t) \notag \\
    \braket{\delta A}(t) &= -i \int_{t_0}^t ds f(s) \Tr'\left\{ \rho_0 [A(t), B(s)] \right\}
\end{align}
where $\rho_0 = \rho(t_0)$ denotes the initial state density matrix and the time-dependence of the operators is governed by the unperturbed Hamiltonian, e.g., $A(t) = e^{ i H_{\rm nh} t} A e^{-i H_{\rm nh} t}$. Note that the use of biorthogonal expectation values is essential to derive these equations which are formally identical to conventional linear-response theory, but allow $H_{\rm nh}$ to be non-hermitian.

The next step is to express $\braket{\delta A}(t)$ using a Lehmann spectral representation based on the biorthogonal completeness relation. Then, assuming an oscillatory perturbation $f(t) = e^{ -i \omega_0 t}$ and using the time evolution of the left and right eigenstates,
\begin{align}
    e^{-i H_0 t} \ket{\phi_m} &= e^{ -i \epsilon_m t} \ket{\phi_m} \notag \\
    \bra{\tphi_m} e^{i H_0 t} &= e^{i \epsilon_m t} \bra{\tphi_m}
\end{align}
makes it possible to perform the time integral and one finds
\begin{align}
    \braket{\delta A}(t)
&=
    \frac{e^{-i \omega_0 t}}{Z} \sum_{m,n}  \bra{\tphi_m} A  \ket{\phi_n} \bra{\tphi_n} B \ket{\phi_m} \frac{e^{ -\beta \epsilon_m}- e^{ -\beta \epsilon_n}}{\omega_0 + \epsilon_m - \epsilon_n} \left( 1 - e^{ i (\omega_0 + \epsilon_m - \epsilon_n) t} \right)
\end{align}
where we have set $t_0 = 0$ for simplicity. Specializing for the case of the Hall conductance, one chooses $A = j_x$ and $B = j_y$ and arrives at Eq.~(2) of the main text.

\section{DC Hall conductance formula for general non-Hermitian Chern insulators}\label{App:methode}

In this section we will outline the method to derive the DC Hall conductance of non-Hermitian Chern insulators of the form
\begin{equation}\label{eq:app1}
 H_{\rm nh}(\vec{k})=[\vec{k}\cdot\bsigma+m\sigma_z]-(\mu+i\Gamma_0)\sigma_0-i\vec{\Gamma}\cdot\bsigma.
\end{equation}
This Hamiltonian is to be interpreted as the quasiparticle Hamiltonian describing a system with gain and loss. In the case of a self-energy being reduced to a non-Hermitian description, the non-Hermitian Hamiltonian is defined based on the inverse \emph{retarded} Green's function. We proceed inversely and postulate a Matsubara Green's function formulation corresponding to the biorthogonal linear response in a non-Hermitian Hamiltonian. In this process, the form of the self-energy needs to be chosen carefully. In the usual case, the real-time formalism is connected to the imaginary time formalism via the replacement $i\omega_n\to\omega \pm i 0^ +$ (for retarded and advanced Green's functions), and the self-energy is typically of the form $\Sigma(i\omega_n)\sim\text{sgn}(\omega_n)$. The sign function has the analytic contination $\text{sgn}(\omega_n) \to \text{sgn}(\eta)$ for $i\omega_n\to\omega+i\eta$ with $\eta\to0$, and can be interpreted as ensuring that both particle and hole excitations decay. We are interested in extending the scenario of loss, which includes the special case of a self-energy encoding a finite lifetime, to both loss and gain, and therefore define the Matsubara Green's function as
\begin{equation}
\mathcal{G} (\vec{k},i\omega) = \frac{1}{i\omega-[\vec{k}\cdot\bsigma+m\sigma_z]+[\mu+i\Gamma_0(i\omega)]\sigma_0+i\vec{\Gamma}(i\omega)\cdot\bsigma},\quad \Gamma_\mu(i\omega) := \Gamma_\mu\text{sgn}\left(\Im i\omega\right).
\end{equation}
From now on, we use the notation $z=i\omega$. Using $\vec{d}=\vec{k}-i\vec{\Gamma}$, we find
\begin{align}
\mathcal{G} (z) =& \frac{1}{ z+\mu+i\Gamma_0(z) -\vec{d}(z)\cdot\bsigma} = \frac{[z+\mu+i\Gamma_0(z)]+\vec{d}(z)\cdot\bsigma}{[z+\mu+i\Gamma_0(z)]^2-\vec{d}(z)\cdot \vec{d}(z)}= \sum_\alpha \frac{P_\alpha(z)}{z+\mu+i\Gamma_0(z) -d_{\alpha}(z)}
\end{align}
with $d_\pm(z) = \pm d(z)$ and
\begin{equation}
P_{\alpha}(\vec{k},z) = \frac{1}{2}\left(1+\frac{\vec{d}(\vec{k},z)\cdot\bsigma}{d_\alpha(\vec{k},z)}\right),\qquad d_\alpha(\vec{k},z) = \alpha\sqrt{(k_x-i\Gamma_x(z))^2+(k_y-i\Gamma_y(z))^2+(m-i\Gamma_z(z))^2}.
\end{equation}

\subsection{General form of the Hall conductance}

The Green's function are now expressed in terms of their spectral decomposition, whereas the derivatives of inverse Green's functions with respect to the momenta are just Pauli matrices. The derivative with respect to the Matsubara frequency is special because it involves the derivative of the sign function. (Anti)-symmetrizing  the Ishikawa-Matsuyama formula in units of $\hbar=e=1$ w.r.t $x$ and $y$, the Hall conductance reads:
\begin{align}
\sigma_{xy} =   \int\frac{d^2k}{(2\pi)^2}\int_{-i\infty}^{i\infty}\frac{dz}{2\pi}\frac{1}{2}&\Bigg\{ \Tr\Big[ \sigma_x P_{\alpha}(\vec{k},z)\sigma_y P_{\beta}(\vec{k},z)\left(\partial_z\mathcal{G}^{-1}(\vec{k},z)\right)P_{\gamma}(\vec{k},z)\Big]\nonumber-\Tr\Big[\sigma_y P_{\alpha}(\vec{k},z)\sigma_x P_{\beta}(\vec{k},z) \left(\partial_z\mathcal{G}^{-1}(\vec{k},z)\right)P_{\gamma}(\vec{k},z)\Big]\Bigg\}\nonumber\\
 &\times \frac{1}{z+\mu+i\Gamma_0(z)-d_\alpha(\vec{k},z)}\frac{1}{z+\mu+i\Gamma_0(z)-d_\beta(\vec{k},z)}\frac{1}{z+\mu+i\Gamma_0(z)-d_\gamma(\vec{k},z)}.
\end{align}
The inverse Green's function reads $\mathcal{G}^{-1}(\vec{k},z)=[z+i\Gamma_0(z)+i\vec{\Gamma}(z)\cdot\bsigma+(\cdots)]$ where $(\cdots)$ are frequency independent terms. The derivative of the sign function is given by $\partial_z \text{sgn}(z) = 2\delta(z)$ and hence the derivative of the inverse Green's function with respect to the Matsubara frequency gives rise to two contributions:
\begin{equation}\label{eq:deriv}
\partial_z\mathcal{G}^{-1}(\vec{k},z) = 1 + 2i\delta(z)\left(\Gamma_0+\vec{\Gamma}\cdot\bsigma\right).
\end{equation}
The constant term $1$ gives rise to Fermi sea contributions since it does not affect the integration over the full frequency range: it contains contribution from all the occupied states. The second term involving the delta function only gives rise to contributions for $z=0$, that is, from the Fermi surface.
We can group the Fermi surface and Fermi sea contributions, and split the DC Hall conductance into
\begin{equation}
\sigma_{xy} = \sigma^{(\rm sea)}_{xy}+\sigma^{(\rm surface)}_{xy},
\end{equation}
where $\sigma^{(\rm surface)}_{xy}$ is the Fermi surface contribution and $\sigma^{(\rm sea)}_{xy}$ the (usual) Fermi sea contribution. In the absence of non-Hermitian terms, the Fermi surface term vanishes and $ \sigma^{(\rm sea)}_{xy}$ is the quantized Hall conductance.

\subsection{Fermi surface contributions}\label{App:Fsurface}
Using $\text{sgn}(0)=0$, the Fermi surface contribution can be written as
\begin{align}
\sigma^{(\rm surface)}_{xy} =&   i\int\frac{d^2k}{(2\pi)^2}\frac{1}{2\pi}\Bigg\{\Tr\Big[ \sigma_x P_{\alpha}(\vec{k})\sigma_y P_{\beta}(\vec{k})\left( \Gamma_0\sigma_0+ \vec{\Gamma}\cdot\bsigma\right)P_{\gamma}(\vec{k})\Big] - \Tr\Big[\sigma_y P_{\alpha}(\vec{k})\sigma_x P_{\beta}(\vec{k}))\left( \Gamma_0\sigma_0+ \vec{\Gamma}\cdot\bsigma\right)P_{\gamma}(\vec{k})\Big]\Bigg\}\nonumber\\ &  \frac{1}{\mu-d_\alpha(\vec{k})}\frac{1}{\mu-d_\beta(\vec{k})}\frac{1}{\mu-d_\gamma(\vec{k})},
\end{align}
where the projector now takes on the simpler form
\begin{equation}
P_{\alpha}(k) = \frac{1}{2}\left(1+\frac{\vec{d}\cdot\bsigma}{d_{\alpha}(\vec{k)}} \right),\quad \vec{d}(\vec{k}) = (k_x,k_y,m),\quad d_{\alpha}(\vec{k})= \alpha\sqrt{k^2+m^2}.
\end{equation}
In the trace, contributions from $\Gamma_x\sigma_x$ and $\Gamma_y\sigma_y$ are zero due to the anti-symmetry of the conductance tensor. Hence, only terms $\propto$ $\Gamma_0$ and $\Gamma_z$ will contribute to the trace.
We can thus write $\sigma^{(\rm surface)}_{xy}$ as the sum of two terms,
\begin{align}
\sigma^{(\rm surface)}_{xy,a} =   i\Gamma_0\int\frac{d^2k}{(2\pi)^2}\frac{1}{2\pi} \Bigg\{\Tr\Big[ \sigma_x P_{\alpha}(\vec{k})\sigma_y P_{\beta}(\vec{k}) P_{\gamma}(\vec{k})\Big] -  \Tr\Big[\sigma_y P_{\alpha}(\vec{k})\sigma_x P_{\beta}(\vec{k})) P_{\gamma}(\vec{k})\Big]\Bigg\} \frac{1}{\mu-d_\alpha(\vec{k})}\frac{1}{\mu-d_\beta(\vec{k})}\frac{1}{\mu-d_\gamma(\vec{k})},
\end{align}
and
\begin{align}
\sigma^{(\rm surface)}_{xy,b} = i\Gamma_z\int\frac{d^2k}{(2\pi)^2}\frac{1}{2\pi} \Bigg\{\Tr\Big[ \sigma_x P_{\alpha}(\vec{k})\sigma_y P_{\beta}(\vec{k})\sigma_z P_{\gamma}(\vec{k})\Big]- \Tr\Big[\sigma_y P_{\alpha}(\vec{k})\sigma_x P_{\beta}(\vec{k}))\sigma_z P_{\gamma}(\vec{k})\Big]\Bigg\}\frac{1}{\mu-d_\alpha(\vec{k})}\frac{1}{\mu-d_\beta(\vec{k})}\frac{1}{\mu-d_\gamma(\vec{k})}.
\end{align}
Performing the trace on both expressions, we find
\begin{align}
\sigma^{(\rm surface)}_{xy,a} =&  -\frac{4}{2\pi}\Gamma_0\int\frac{d^2k}{(2\pi)^2} \frac{m}{\left[(k^2+m^2)-\mu^2\right]^2}
=  -\frac{1}{(2\pi)^2}\frac{2m\Gamma_0}{m^2-\mu^2}, \nonumber \\
\sigma^{(\rm surface)}_{xy,b} =&   -\frac{4}{2\pi}\Gamma_z\int\frac{d^2k}{(2\pi)^2} \frac{\mu}{\left((k^2+m^2)-\mu^2\right)^2}
=    -\frac{1}{(2\pi)^2}\frac{2\mu \Gamma_z}{m^2-\mu^2}.
\end{align}
Hence, the result of this contribution reads, retrieving units
\begin{equation}\label{eq:Fsur}
\sigma^{(\rm surface)}_{xy} = -\frac{1}{ \pi }\frac{e^2}{h}\frac{\mu \Gamma_z+m \Gamma_0}{m^2-\mu^2}.
\end{equation}

\subsection{Fermi sea contributions}\label{App:Fsea}
Evaluating the trace, the Fermi sea contribution can be brought to the form
\begin{align}
\sigma^{(\rm sea)}_{xy} =   \int\frac{d^2k}{(2\pi)^2}\int_{-i\infty}^{i\infty}\frac{dz}{2\pi}\frac{1}{2} \frac{4 i [m-i\Gamma_z(z)]}{[d^2(\vec{k},z)-(z+i\Gamma_0(z)+\mu)^2]^2}.
\end{align}
Defining $\varepsilon_\pm(\vec{k}) =  \sqrt{(k_x\mp i\Gamma_x)^2+(k_y \mp i\Gamma_y )^2+(m\mp i\Gamma_z)^2}$, we obtain
\begin{align} \label{eq:Fsea}
\sigma^{(\rm sea)}_{xy} &= 2i \int\frac{d^2k}{(2\pi)^2}\int_{i0^+}^{+i\infty}\frac{dz}{2\pi}\left\{ \frac{ (m+i\Gamma_z)}{[\varepsilon^2_-(\vec{k})-(z+i\Gamma_0-\mu)^2 ]^2}+ \frac{(m-i\Gamma_z)}{[\varepsilon^2_+(\vec{k})-(z+i\Gamma_0+\mu)^2 ]^2}\right\}.
\end{align}
The two terms can be interpreted as contributions from the conduction band and valence band of the system. The information about dissipation and the interaction is contained in the $\varepsilon_\pm(\vec{k})$ terms. Specializing to the case $\Gamma_y\neq0$ but $\Gamma_x=0=\Gamma_z$, and with $z=i\omega$, we obtain:
\begin{align}
\sigma^{(\rm sea)}_{xy}
&=  -2 m\int\frac{dk_x}{(2\pi)^2} \int_{0}^{\infty}\frac{d\omega}{2\pi}\sum_{\alpha=\pm} \int dk_y  \frac{1}{ \left[(k_y-i\Gamma_y)^2+ [k_x^2+m^2 +(\alpha i\mu+\Gamma_0+\omega)^2 ] \right]^2} .
\end{align}
The $k_y$ integrals can be evaluated using the residue theorem, which yields
\begin{align}
\sigma^{(\rm sea)}_{xy}=&   - m\int\frac{dk_x}{(2\pi)^2} \int_{0}^{\infty}d\omega\Re \frac{1}{\sqrt{k_x^2+m^2 +(i\mu+\Gamma_0+\omega)^2}^3}\Theta\left[-\mathcal{A}(k_x,\omega+\Gamma_0)\right].
\end{align}
Using $\Theta(-x)=1-\Theta(x)$, we decompose $\sigma^{(\rm sea)}_{xy}$ as a contribution $\tilde{\sigma}^{(\rm sea)}_{xy}$ without phase-space restriction minus a contribution which corresponds to the conductance contribution $\sigma^{(3/2)}$ discussed in the main text,
\begin{equation}\label{eq:Fseatotal}
\sigma^{(\rm sea)}_{xy} = \tilde{\sigma}^{(\rm sea )}_{xy}+\sigma^{(\rm 3/2)}_{xy},
\end{equation}
where
\begin{equation}\label{eq:Fseatilde}
\tilde{\sigma}^{(\rm sea )}_{xy} =   - \frac{m}{|m|}\frac{e^2}{h}\frac{1}{2\pi}\left[\frac{\pi}{2}-\arctan\left( \frac{\Gamma_0^2+\mu^2-m^2}{2|m|\Gamma_0}\right)\right],
\end{equation}
as already derived in Refs.~\cite{Hirsbrunner2019,Philip2018} and $\sigma^{(\rm 3/2)}_{xy}$ is given, after a shift of variable $\omega + \Gamma_0 \to \omega$ in the main text.

\end{document}